\documentclass[reprint,amsmath,amssymb,aps,prd]{revtex4-2}
\pdfoutput=1
\usepackage{filecontents}
\usepackage{graphicx}
\usepackage{dcolumn}
\usepackage{bm}
\usepackage[usenames,dvipsnames]{color}
\usepackage{longtable}
\usepackage{tensor}
\usepackage{epsfig}
\usepackage{amssymb}
\usepackage{amsmath}
\usepackage{verbatim}
\usepackage{xspace}
\usepackage{aas_macros}
\include{jdefs}
\usepackage{graphics,hyperref}
\newcommand{\wt}{{\hspace{-0.05cm}\scalebox{0.65}{\textit{T}}}}

\begin{document}
\title{The phase diagram of Einstein-Weyl gravity}
\author{S.~Silveravalle${}^{1,2}$ and A.~Zuccotti${}^{3}$}

\affiliation{
\mbox{${}^1$ Universit\`a degli Studi di Trento,\\Via Sommarive, 14, IT-38123, Trento, Italy} \\
\mbox{${}^2$ INFN - TIFPA,\\Via Sommarive, 14, IT-38123, Trento, Italy} \\
\mbox{${}^3$ Ghent University,\\ Technologiepark-Zwijnaarde 126, Be-9052 Gent, Belgium}
}

\begin{abstract}
Thanks to their interpretation as first order correction of General Relativity at high energies, quadratic theories of gravity gained much attention in recent times. Particular attention has been drawn to the Einstein-Weyl theory, where the addition of the squared Weyl tensor to the action opens the possibility of having non-Schwarzschild black holes in the classical spectrum of the theory. Static and spherically symmetric solutions of this theory have been studied and classified in terms of their small scales behaviour; however, a classification of these solutions in terms of the asymptotic gravitational field is still lacking. In this paper we address this point and present a phase diagram of the theory, where the different types of solutions are shown in terms of their mass and the strength of a Yukawa-like correction to the gravitational field. In particular we will show that, in the case of compact stars, different equations of state imply different Yukawa corrections to the gravitational potential, with possible phenomenological implications.
\end{abstract}

\maketitle

\section{Introduction}
General Relativity is one of the most successful theory of last century, however there is still no general consensus on how gravity should be described at the quantum level. 
It is known that the Einstein-Hilbert action of General Relativity is not renormalizable within standard perturbative methods, and modifications of such action are expected at high energies.
The study of modified gravity theories is largely used to address the high energy limit of gravity while preserving General Relativity at lower energies.  \\
The first corrections expected are quadratic terms in the curvatures \cite{tHooft:1974toh}, which in 4 dimensions modify the action as
\begin{equation}\label{sqg}
		S_{QG} = \int{ d^4 x \sqrt{-g} [\gamma R+\beta R^2 -\alpha C_{\mu \nu \rho \sigma}C^{\mu \nu \rho \sigma}+\mathcal{L}_m]}.
\end{equation} 
Indeed such action has a long history: in \cite{Stelle:1976gc} it is proved that the action (\ref{sqg}) in the vacuum is perturbatively renormalizable, in \citep{Zwiebach:1985uq} it appears in the low energy limit of string theory, and recently it has emerged in the framework of the renormalization group flow \cite{machado,Benedetti:2009rx,Hamada:2017rvn}, Asymptotically Safe program \cite{Percacci:2017fkn,Reuter:2019byg} and 
fakeons theory \cite{anselmi17,Anselmi:2018ibi}.
The physical content of (\ref{sqg}) can be resumed in the standard massless graviton, plus a massive scalar mediator and a massive spin two ghost mediator, which at the quantum level implies the loss of unitarity. While various authors proposed solutions to the ghost problem, in this work we want to focus on the classical content of the quadratic action. In particular we are interested in studying the case of static spherically symmetric solutions without a cosmological constant in order to describe the gravitational field of isolated objects. Despite the classical solutions of (\ref{sqg}) have been largely investigated in recent works \cite{Podolsky:2018pfe,Podolsky:2019gro,Saueressig:2021wam,Lu:2015psa,Goldstein:2017rxn}, given the non linear nature of the field equations, an exact form of the general solution is lacking, and numerical methods have to be used in order to understand the link between the asymptotic field and the physical nature of the solution.
In the following we consider the Einstein-Weyl action
\begin{equation}
    \label{actionew}
   \mathcal{S_{EW}}=\int\mathrm{d}^4x\,\sqrt{-g}\,\left[\gamma\,R-\alpha\,C^{\mu\nu\rho\sigma}C_{\mu\nu\rho\sigma}+\mathcal{L}_m\right],
\end{equation}
i.e. the quadratic theory restricted to the $\beta=0$ case. Both the theories defined in (\ref{sqg}) and (\ref{actionew}) present a large variety of solution families in addition to the Schwarzschild one, but while the effect of the $R^2$ term has been largely studied in astrophysical and cosmological contexts \cite{Starobinsky:1980te,Orellana:2013gn,Astashenok:2013vza,Sbisa:2019mae}, the $C^2$ term received relatively less attention until recent times. The Einstein-Weyl theory indeed gives interesting insight on the physical content of quadratic gravity. It is proved that the solution space of Einstein-Weyl gravity coincides with the one of (\ref{sqg}) with the constraint $R=0$, and a no-hair theorem presented in \cite{Nelson:2010ig}, and corrected in \cite{Lu:2015psa}, states that under certain conditions the Ricci scalar must vanish. As main consequence this theorem implies that all the asymptotically flat black hole solutions of quadratic gravity are present only in the Einstein-Weyl restriction. Moreover, the results in \cite{Lu:2015psa} show that, together with black holes, the main new families of solutions appearing in quadratic gravity, also appear in the Einstein-Weyl theory. Finally, it has also been shown that the $C^2$ has a much stronger effect than the $R^2$ one in the properties of compact stars \cite{Bonanno:2021zoy}.\\ 
While attempts to link the different families of solutions to the asymptotic field have been made in the full theory \cite{Daas:2022iid,Silveravalle:2022lid}, the presence of two massive mediators in the full quadratic case brings numerical issues when integrating the asymptotic field, namely some non-linear terms can be larger than linear ones, which affects the physical results obtained. Such numerical instabilities have not been encountered in the Einstein-Weyl theory, so we found sensible to restrict our results to this case where the numerical procedure is much more reliable.
In this paper we use the analytical approximation found in previous works, together with numerical shooting techniques, in order to show the complete solution space of the Einstein-Weyl gravity in form of phase diagram of the theory; to the best of our knowledge this is the first time the link between the asymptotic field and the families of solutions is shown with this level of robustness.
Building the phase diagram of the theory is a crucial step to understand the physical content since it allows to connect the type of solution with the observables at large distances. In particular the results encoded in the phase diagram gives new insight on the structure of the solution space of quadratic gravity, showing that some of the new families of vacuum solution expected in previous work appear together with Schwarzschild black holes for all positive mass values, while others are confined in a finite mass interval.\\
In what follows we first recall the analytical approximation needed as boundary condition and we describe the numerical method used to integrate the field equations. Then we list the vacuum solution families encountered in the solution space, the behavior of their gravitational potential, together with their causal structure. We present the phase diagram of the theory, in which is indicated the type of solution in function of the gravitational field at large distances. The case of a self-gravitating perfect fluid is studied, showing the corresponding gravitational field and mass-radius relation compared to their General Relativity counterpart.
 
\section{Equations of motion, analytical approximations and numerical integration}

The equations of motion of Einstein-Weyl gravity in tensorial form are
\begin{equation}
\begin{split}
    \label{eomew1}
    \mathcal{H}_{\mu\nu}&=\gamma \left(R_{\mu\nu}-\frac{1}{2}R\,g_{\mu\nu}\right)+\\
    &-4\,\alpha\left(\nabla^\rho\nabla^\sigma +\frac{1}{2}\,R^{\rho\sigma} \right)C_{\mu\rho\nu\sigma}=\frac{1}{2}T_{\mu\nu},
    \end{split}
\end{equation}
with the trace being
\begin{equation}\label{traceeom}
    \mathcal{H}^{\mu}_{\ \mu}=\gamma R=\frac{1}{2}T^{\mu}_{\ \mu}.
\end{equation}
Given the condition $\nabla^{\mu}\mathcal{H}_{\mu\nu}=0$ and $\mathcal{H}_{\phi\phi}=\sin^2\theta\  \mathcal{H}_{\theta\theta}$, only two of these equations are independent.
As the ansatz for the static spherically symmetric metric we choose the one with Schwarzschild coordinates
\begin{equation}
    ds^2=-h(r)dt^2+\frac{dr^2}{f(r)}+r^2d\Omega^2.
\end{equation}
The equations of motion result equivalent to a system of two second order ordinary differential equations in $h(r)$, $f(r)$, as it is shown in \citep{Lu:2015psa,Bonanno:2019rsq}, corresponding to
\begin{equation}
    \begin{split}\label{eomew2}
\tensor{\mathcal{H}}{^\mu_\mu}&=\gamma R=\frac{1}{2}\,\tensor{T}{^\mu_\mu},\\
\mathcal{H}_{rr}&+A(r)\,\partial_r\left(\tensor{\mathcal{H}}{^\mu_\mu}-\frac{1}{2}\,\tensor{T}{^\mu_\mu}\right)+\\ &+B(r)\,\left(\tensor{\mathcal{H}}{^\mu_\mu}-\frac{1}{2}\,\tensor{T}{^\mu_\mu}\right)^2+\\
&+C(r)\left(\tensor{\mathcal{H}}{^\mu_\mu}-\frac{1}{2}\,\tensor{T}{^\mu_\mu}\right)=\frac{1}{2}\,T_{rr},
\end{split}
\end{equation}
where $A(r)$, $B(r)$ and $C(r)$ are combinations of the coupling $\alpha$, the metric functions $h(r)$, $f(r)$ and their first derivatives.
We show these equations explicitly in the vacuum case
\begin{equation}\label{eomvacuum}
\begin{split}
&4 h(r)^2 \big(r f'(r)+f(r)-1\big)-r^2 f(r) h'(r)^2\\
&+r h(r) \big(r f'(r) h'(r)
+2 f(r) \big(r h''(r)+2 h'(r)\big)\big)=0,\\[0.3cm]
&\alpha  r^2 f(r) h(r) \big(r f'(r)+3 f(r)\big) h'(r)^2\\
&+2 r^2 f(r) h(r)^2 h'(r) \big(\alpha  r f''(r)+\alpha  f'(r)-\gamma  r\big)\\
&+h(r)^3 \big(r \big(3 \alpha  r f'(r)^2-4 \alpha  f'(r)+2
   \gamma  r\big)\\
&-2 f(r) \big(4 \alpha +2 \alpha  r^2 f''(r)-2 \alpha  r f'(r)+\gamma  r^2\big)\\
&+8 \alpha  f(r)^2\big)-\alpha  r^3 f(r)^2 h'(r)^3=0.
\end{split}
\end{equation}
To our knowledge, no analytical solutions of such system have been found, with the exception of the ones present in General Relativity \textit{i.e.} the Minkowski and the Schwarzschild spacetime. Numerical methods and analytical approximations have to be used in order to study the complete solution space.   

\subsection{Linearized solutions at large distance}

Since we are interested in studying isolated objects without a cosmological constant, that is we look for asymptotically flat solutions, we can describe the metric at large distances using the weak field limit. As described in \citep{Stelle:1977ry,Lu:2015psa,Bonanno:2019rsq}, we write the functions $h(r)$ and $f(r)$ as
\begin{equation}
h(r)=1+\epsilon\,V(r),\qquad f(r)=1+\epsilon\,W(r),
\end{equation}
and solve (\ref{eomew1}) at linear order in $\epsilon$. 

\begin{table*}[t]
\centering
\caption{Families of solutions around finite and zero radii in Einstein-Weyl gravity, as shown in \cite{Zuccotti:2022}}
\begin{tabular}{c c c}\hline\hline\\[-0.27cm]
\phantom{aaaaaaaaa}Family\phantom{aaaaaaaaa} & \phantom{aaaaaaaa}$\mathrm{N}^{\mathrm{o}}$ of free parameters\phantom{aaaaaaaa} & \phantom{aaaaaaaaaa}Interpretation\phantom{aaaaaaaaaa}\\[0.05cm] \hline\\[-0.25cm]
$(0,0)_0^1$  & $2\,(\to 0)$ & Regular solution/True vacuum\\ 
$(-1,-1)_0^1$ & $3\,(\to 1)$ & Naked singularity/Schwarzschild interior\\ 
$(-2,2)_0^1$ & $4\,(\to 2)$ & Bachian singularity/Holdom star\\ 
$(0,0)_{r_0}^1$ & $4\,(\to 2)$ & Regular metric\\
$(1,1)_{r_0}^1$ & $3\,(\to 1)$ & Black hole\\
$(1,0)_{r_0}^1$ & $2\,(\to 0)$ & Symmetric wormhole\\
$(1,0)_{r_0}^2$  & $4\,(\to 2)$ & Non-symmetric wormhole\\
$(4/3,0)_{r_0}^3$ & $3\,(\to 1)$ & Not known\\[0.1cm]
\hline \hline
\end{tabular}
\label{tabfa}
\end{table*}

When imposing asymptotic flatness and fixing $h(r)\to 1$ as $r\to+\infty$, it is possible to show that the solutions result to be
\begin{equation}\begin{split} \label{soluzioni linearizzate}
h(r)=\, &1-\frac{2\,M}{r}+2S^-_2 \frac{e^{-m_2\, r}}{r}, \\
f(r)=\, &1-\frac{2\,M}{r}+S^-_2 \frac{e^{-m_2\, r}}{r}(1+m_2\, r) ,
\end{split}\end{equation}
with $m_2^2=\frac{\gamma}{2\alpha}$ being the mass of the spin-two ghost, and $M$ the ADM mass in Planck units. We note that in the non-relativistic limit the gravitational potential will have a Yukawa correction, as expected for a massive mediator, and therefore we will refer to the parameter $S_2^-$ as the Yukawa charge.

\subsection{Series expansion at finite radii}

At finite radius the solution can be found within series expansions by using a generalized Frobenius method. The metric functions can be expanded as
\begin{equation}
\begin{split}
h(r)&=\left(r-r_0\right)^t\!\left[\displaystyle\sum_{n=0}^N h_{t+\frac{n}{\Delta}}\left(r-r_0\right)^{\frac{n}{\Delta}}\!+O\!\left(\left(r-r_0\right)^{\frac{N\!+\!1}{\Delta}}\right)\!\right]\!,\\
f(r)&=\left(r-r_0\right)^s\!\left[\displaystyle\sum_{n=0}^N f_{s+\frac{n}{\Delta}}\left(r-r_0\right)^{\frac{n}{\Delta}}\!+\!O\left(\left(r-r_0\right)^{\frac{N\!+\!1}{\Delta}}\right)\!\right]\!,
\end{split}
\end{equation}
and it is possible to classify the solutions as $(s,t)_{r_0}^\Delta$. In \citep{Lu:2015psa,Podolsky:2019gro} the families of solution allowed have been exhaustively studied. The complete list of such families is shown in Table \ref{tabfa}, where in the second column is reported the number of free parameters after imposing asymptotic flatness and a specific time parameterization. 
We specify that we used a different notation from the one in \citep{Lu:2015psa,Podolsky:2019gro}, in particular the different sign for the exponent $s$ due to the metric ansatz in terms of the function $f(r)$ instead of $A(r)=1/f(r)$ for the families around $r_0=0$.

\subsection{Numerical integration and Shooting method}

In what follows the numerical results discussed are expressed in unit of $m_2=\sqrt{\frac{\gamma}{2\alpha}}$. We used $\frac{\gamma}{2\alpha}=1$ in the numerical code, together with $G=1$ when evaluating the linerized solution (\ref{soluzioni linearizzate}), in order to obtain adimensional results. Physical units are restored for non-vacuum solutions, in order to have a comparison with the solutions of General Relativity.

The general asymptotically flat solutions in the vacuum have been studied by numerically integrating the e.o.m. (\ref{eomvacuum}) with the linearized solution (\ref{soluzioni linearizzate}) as boundary conditions.
To integrate the equations of motion we used the Adaptive Stepsize Runge-Kutta integrator DO2PDF implemented by the \texttt{NAG} group (see \url{https://www.nag.com} for details) with a tolerance of $10^{-12}$. The same analysis was independently carried out with the NdSolve function from \texttt{Wolfram Mathematica} by using an adaptive stepsize method which switches between a midpoint and an implicit Euler methods, with a tolerance of $10^{-14}$. The results obtained are in agreement with the previous one of the \texttt{NAG} code. 
The solutions have been classified in terms of the behaviour of the two functions
\begin{equation}
    \begin{split}
        \label{chis}
        \chi_h(r) =r\frac{h'(r)}{h(r)},\qquad        \chi_f(r) =r\frac{f'(r)}{f(r)}
    \end{split}
\end{equation}
evaluated close to the origin which are expected to match the values $s,t$ of the corresponding family in Tab. \ref{tabfa}. In particular we evaluated the $\chi_{f/h}(r)$ functions at the radius $r_O=10^{-2}$. Wormholes and black-holes are found anytime it is not possible to integrate the e.o.m. to the origin.
In those cases we performed a more detailed analysis by solving a boundary value problem (BVP): as external boundary conditions we used again the linearized solutions (\ref{soluzioni linearizzate}), while as internal conditions we used the specific series expansion of the wormhole/black hole family. As large radius we opted for a value $r_\infty=18$ in order to have Yukawa corrections bigger than the tolerance threshold. The wormhole/black hole series has been evaluated at $r_{T/h}+10^{-3}$ where $r_{T/h}$ is either the radius of the wormhole throat or the black hole horizon.
The BVP is solved by implementing the Shooting Method to a fitting point, where a globally convergent Broyden's method with a tolerance of $10^{-6}$ in the \texttt{NAG} code and $10^{-4}$ in the \texttt{Wolfram} one is used in order to obtain continuity of the metric functions and their derivatives. The precise value of the fitting radius does not affect the accuracy of the shooting method, but is important for obtaining convergence efficiently. In particular a fitting radius $r_{fit}=r_h+0.3$ for black holes and $r_{fit}=r_T+0.05$ has been found optimal. 
\begin{figure*}[t]
\centering
\includegraphics[width=\textwidth]{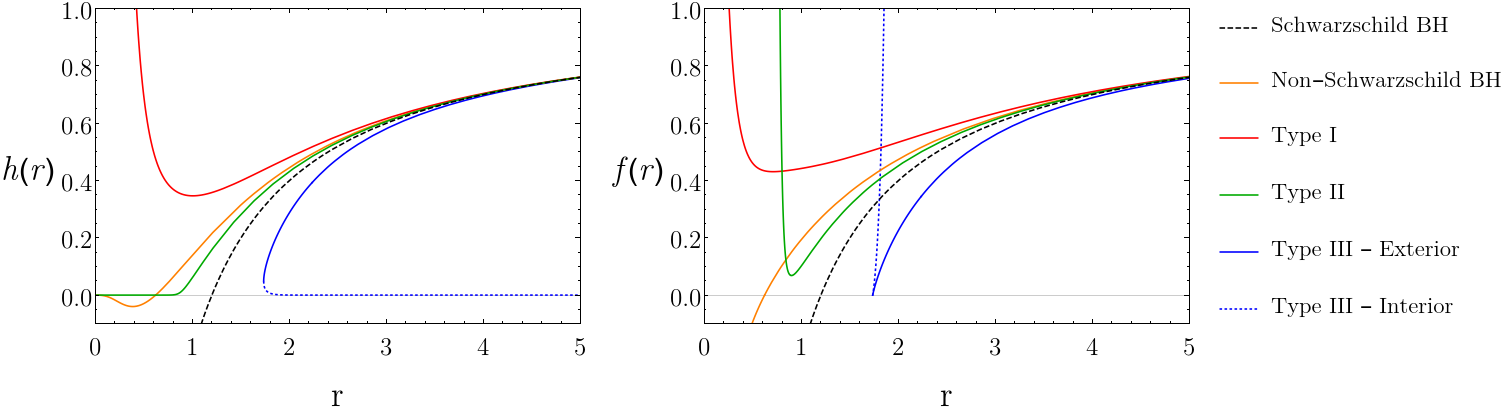}
\caption{Vacuum solutions of Einstein-Weyl gravity with mass $M=0.6$. The Schwarzschild BH in dashed black has $S_2^-=0$, the non-Schwarzschild BH in orange has $S_2^-=0.101$, the type I solution in red has $S_2^-=0.2$, the type II solution in green has $S_2^-=0.075$, and the type III solution in dotted and solid blue has $S_2^-=-0.2$.}
\label{solutions}
\end{figure*}
As we will see in the following section, the Shooting Method allowed us to determine the precise position of black holes and wormholes on the phase diagram, as well as to continue the integration behind the horizon/throat radius. Moreover it has been possible to extract the relevant physical properties of the solutions around $r_{T/h}$ in function of the gravitational parameters at large distances.

\section{Solutions of Einstein-Weyl gravity}

In this section we summarize the main properties of the different families of solutions. We recap the metric behaviour shown in previous works, as in \cite{Lu:2015psa,Bonanno:2019rsq,Zuccotti:2022,Saueressig:2021wam,Holdom:2016nek,Bonanno:2021zoy}, and briefly sketch their physical behaviour.

\subsection{Type I: $(-1,-1)_0^1$ solutions}

The first class of solution found, namely type I solutions, are characterized by values of $(\chi_{f}(r),\chi_{h}(r))$ between $-0.8$ and $-1.4$ for both the metric functions. 
This suggests that such solutions should be given by the $(-1,-1)_0^1$ family i.e.
\begin{equation}\begin{split}
ds^2=&-\Bigl(\frac{h_{-1}}{r}+h_0+O(r)\Bigr)dt^2\\
&+\frac{1}{\frac{f_{-1}}{r}+f_0+O(r)}dr^2+r^2 d\Omega^2.
\end{split}\end{equation}
However we have found relevant discrepancies from the expected value $\chi_{f/h}(r)=-1$ even at radii smaller than $r_O=10^{-2}$.
Moreover these solutions result to cover most part of the solution space, while the number of free parameters reported in Tab. \ref{tabfa} suggests that they should occupy a one-dimensional region. These considerations make evident that Type I solutions should actually belong to some correction of the $(-1,-1)_0^1$ family but with the same leading order. In \citep{perkinsthesis} a non-Frobenius family that shares the same leading order of the $(-1,-1)_0^1$ family but with logarithmic corrections and with one additional free parameter has been found in the full quadratic theory, but in Einstein-Weyl gravity none of such family has been discovered. 

Despite the analytical form of such solutions is still unknown, we can extract some physical information from our results. Type I solutions have the same curvature invariants scaling of $(-1,-1)_0^1$
\begin{equation}\begin{split}
    R_{\mu\nu}R^{\mu\nu}&\underset{r\to 0}{\sim} O(r^{-6}),\\
    R_{\mu\nu\rho\sigma}R^{\mu\nu\rho\sigma}&\underset{r\to 0}{\sim} O(r^{-6}),
\end{split}\end{equation}
hence the causal structure of such solutions is the one of a spacetime with naked singularity, since no horizon appears, as it is shown on the left of Fig. \ref{penrose}. In the positive mass region of the solution space, the type I space-time has an attractive gravitational potential at large distances that reaches a minimum and then becomes repulsive around the singular origin, since the temporal component $h(r)$ diverges with a positive sign. 

\subsection{Type II: $(-2,2)_0^1$ solutions}
\begin{figure*}[t]
\centering
\includegraphics[width=\textwidth]{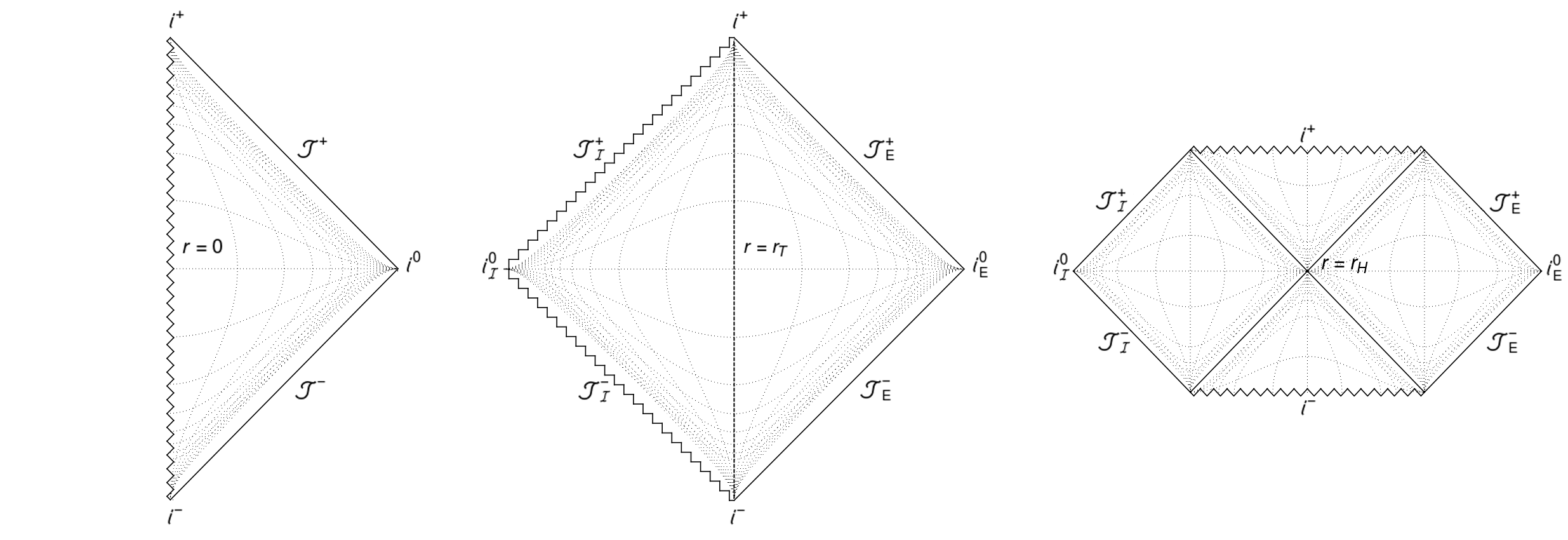}
\caption{Conformal diagrams of a naked singularity (on the left), of a no-sy WH (in the center) and of a black hole (on the right); the dotted lines indicate surfaces of constant time and radius. The conformal diagram of a no-sy WH is taken from \cite{Zuccotti:2022}.}
\label{penrose}
\end{figure*}
Type II solutions are characterized by values of $(\chi_{f}(r),\chi_{B}(r))$ close to $(-2,2)$ around the origin.
This suggests that type II solutions belong to the $(-2,2)_0^1$ family and, in contrast with type I solutions, the area populated by these solutions agrees with the number of free parameters of the $(-2,2)_0^1$ family. Therefore the metric around the origin can be safely be approximated by the expansion
\begin{equation}\begin{split}
ds^2=&-\Bigl(h_2 r^2+O(r^3)\Bigr)dt^2\\
&+\Bigl(\frac{1}{f_{-2}} r^2+O(r^3)\Bigr)dr^2+r^2 d\Omega^2.
\end{split}\end{equation}
These solutions are characterized by a vanishing metric in $r=0$ as it shown in green in Fig. \ref{solutions}, with an attractive singularity characterized by the following invariant
\begin{equation}\begin{split}
    R_{\mu\nu}R^{\mu\nu}&\underset{r\to 0}{\sim} O(r^{-8}),\\
    R_{\mu\nu\rho\sigma}R^{\mu\nu\rho\sigma}&\underset{r\to 0}{\sim} O(r^{-8}).
\end{split}\end{equation}
The origin results to be an infinite redshift point but when solving the geodesic equation it is possible to prove that a distant observer can communicate with the origin in a finite time interval. Hence the causal structure of such solutions is again the one of a naked singularity shown on the left in Fig. \ref{penrose}. In the positive mass region, type II solutions has an attractive gravitational potential in all the space due to the monotonicity of $h(r)$. 
Such solutions have been studied by Holdom \citep{Holdom:2016nek} as possible candidate for the internal solutions of ultra-compact matter sources.

\subsection{Type III: non-symmetric wormholes}

Type III solutions corresponds to non-symmetric wormhole solutions (no-sy WH). The metric around the throat radius is given by the $(1,0)_{r_T}^2$ family
\begin{equation}\begin{split}
ds^2=&-h_0\Bigl(1\pm h_{1/2}(r-r_\wt)^{\frac{1}{2}}+O(r-r_\wt)\Bigr)dt^2\\
&+\frac{1}{f_1 (r-r_\wt)\pm O((r-r_\wt)^{\frac{3}{2}})}dr^2+r^2 d\Omega^2.
\end{split}\end{equation}
The different sign corresponds to the two patches of spacetime connected at the throat.
Indeed, the wormhole nature of these type of solution is manifest after the coordinate transformation $\rho=\pm 2\sqrt{r-r_T}.$ The metric written in terms of $\rho$ is indeed well behaved in $\rho=0$, and geodesics can be smoothly extended from $\rho>0$ to $\rho<0$ \citep{Lu:2015psa}. This type of spacetime consist in two regions both mapped by $r\in [r_\wt,+\infty )$ corresponding to positive or negative $\rho$. In contrast with standard wormhole, the metric is not symmetric under the transformation $\rho \to -\rho$. The peculiar properties of these wormholes have been described in detail in \citep{Zuccotti:2022}. In particular it is shown that a non-asymptotically flat behavior is present in the second patch of the spacetime, which at large radii is given by
\begin{equation}\begin{split}\label{asymptotically vanishing metric}
ds^2=&-C_h\, r^2 \mathrm{e}^{-a\,r}(1+O(\mathrm{e}^{- a\,r}))dt^2\\
&+C_f\, r^2\mathrm{e}^{-a\,r}(1+O(\mathrm{e}^{-a\,r}))dr^2+r^2 d\Omega^2,
\end{split}\end{equation}
where $a$, $C_h$, and $C_f$ are constant positive parameters.
An example of no-sy WH is shown in Fig. \ref{solutions}, with the asymptotically flat patch shown in solid blue, and the asymptotically vanishing in dotted blue.
The behavior (\ref{asymptotically vanishing metric}) is singular for $r\to+\infty$, which actually result to be a region located at a finite proper distance from the throat radius. The metric results asymptotically vanishing for $r\to +\infty$ in the second patch, similarly to the $(-2,2)_0^1$ solutions in the origin. However, by solving the geodesic equation it can be seen that a distant observer can communicate with the singular region only in an infinite amount of time. Therefore, the conformal diagram of such solutions is the one shown on the right in Fig. \ref{penrose} in which the singularity is naked only in its infinite past. 
The gravitational potential of such solutions results always attractive in direction of the singularity, i.e. attractive in direction of the throat in the asymptotically flat side and repulsive in the asymptotically vanishing side.  

\subsection{Black holes}

Black hole solutions, \textit{i.e.} solutions with an horizon, are given by the $(1,1)_{r_h}^1$ expansion around the radius $r_h$
\begin{equation}\begin{split}
ds^2=&-h_1\Bigl(r-r_\wt+O((r-r_\wt)^2)\Bigr)dt^2\\
&+\frac{1}{f_1 (r-r_\wt)+O((r-r_\wt)^2)}dr^2+r^2 d\Omega^2.
\end{split}\end{equation}
As already discussed in \citep{Lu:2015cqa,Goldstein:2017rxn} both Schwarzschild and non-Schwarzschild black holes are present, and in \cite{Bonanno:2019rsq} their metric has been completely characterized.  
These new black holes have different thermodynamical properties than Schwarzschild ones, in particular their energy and entropy decrease as the horizon grows until they reach negative values. The presence of such negative values is related to the ghost nature of the Weyl term. The sign of the Yukawa charge determines different behavior at the origin, as shown in \cite{Bonanno:2019rsq}: negative Yukawa BHs have a divergent metric that goes like type I solutions in the origin, while positive Yukawa ones have a vanishing metric that goes like type II solutions. The casual structure is the same of the Schwarzschild solution since both type of BHs are characterized by one single horizon with an internal spacelike singularity of either $(-1,-1)_0^1$ or $(-2,2)_0^1$ type. 

\subsection{Non-vacuum solutions}

\begin{figure}[ht]
\centering
\includegraphics[width=\columnwidth]{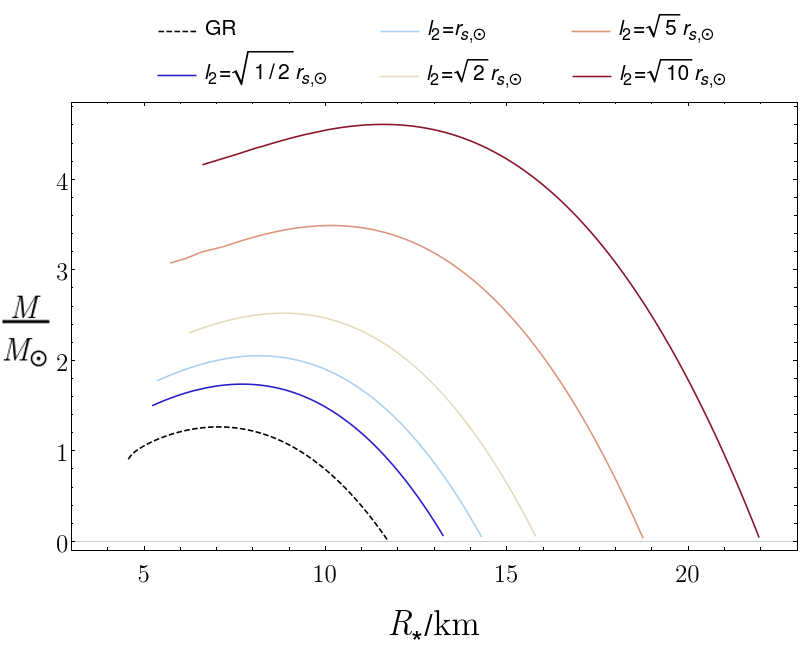}
\caption{Mass-radius relation for polytropic stars in Einstein-Weyl gravity; the equation of state is taken with $\Gamma =2$ and $k_0=6.51185\cdot 10^{-17}\, cm^3/g$ and the scale is fixed in terms of the length unit $l_2=1/m_2$ and the Sun Schwarzschild radius.}
\label{massradius}
\end{figure}

In order to study non-vacuum solutions we consider the stress-energy tensor of a static and isotropic perfect fluid
\begin{equation}
    \label{set}
T_{\mu\nu}=\big(\rho(r)+p(r)\big)\,u_\mu u_\nu+p(r)\,g_{\mu\nu},
\end{equation}
where $u^\mu$ is a unit timelike vector, $\rho(r)$ is the energy density and $p(r)$ is the pressure.
The pseudo conservation equation $\nabla_\mu T^{\mu\nu}=0$ is simplified by the symmetries as 
\begin{equation}
    \label{conset}
p' (r)=-\frac{h' (r)}{2\,h(r)}\big(\rho(r)+p(r)\big),
\end{equation}
and together with an equation of state $p=\mathcal{P}(\rho)$ and the equations (\ref{eomew2}), they form a system of ordinary differential equation in $h(r),\, f(r),\, \rho(r),\, p(r)$. As equations of state we considered polytropes  
\begin{equation}\label{poly}
p(r)=k_0\,\rho(r)^\Gamma,
\end{equation}
with polytropic exponents $\Gamma=2,\, 5/3, 4/3$, and different values of $k_0$; the physical value of $k_0$ is scale dependent, and therefore we simply considered values between $\left[2\cdot 10^{-2}-2\cdot 10^{-1}\right]$ in code units.
In order to integrate the equations we built a shooting code which interpolate between the weak field expansion (\ref{soluzioni linearizzate}) at large distances, and a regular metric in the origin, that is a solution of the $(0,0)_0^1$ family. The integration is performed from the origin to a fitting radius in presence of $\rho(r)$ and $p(r)$, while the external region is integrated in the vacuum. The continuity of the metric function, their derivatives, and of $\rho(r)$ and $p(r)$ is imposed at the fitting radius that, matching the vacuum and non-vacuum integration, is identified as the surface radius; a more detailed explanation can be found in \cite{Bonanno:2021zoy}.\\
As described in \cite{Bonanno:2021zoy}, the compact star solutions found in Einstein-Weyl gravity have no particular difference from the ones found in General Relativity at the level of metric or curvature invariant behaviour. The main effect due to the presence of the quadratic correction is a massive weakening of the gravitational interaction, with the same pressure being able to sustain much more massive stars than in General Relativity. As it can be seen in Fig. \ref{massradius}, where we introduced physical units to have a better comparison with the results from GR, this effect has a strong impact in the mass-radius relation, with an increase both in the maximum mass and radius of stars with the same equation of state. This increase is tightly linked with the effective radius of interaction of the Yukawa particle $l_2=1/m_2=\sqrt{2\alpha/\gamma}$, which however is always smaller than the star surface.

\section{The phase diagram of Einstein-Weyl gravity}

\begin{figure}[ht]
\centering
\includegraphics[width=\columnwidth]{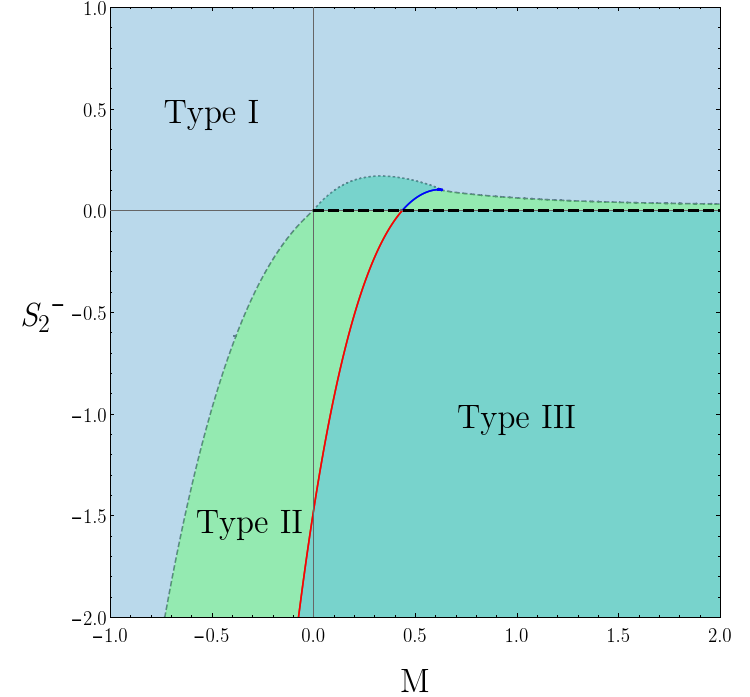}
\caption{Phase diagram of vacuum solutions of Einstein-Weyl gravity; the areas populated by type I, II and III solutions are indicated with three different colors, Schwarzschild black holes are indicated with a black dashed line, while non-Schwarzschild black holes are indicated by the blue and red solid lines. The separation between type I and type II and III solutions are indicated with a dashed and dotted gray lines.}
\label{pd}
\end{figure}

While the existence of type I, type II, no-sy WHs and black hole solutions in Einstein-Weyl gravity was already known, the gap left in the literature that we aim to fill in this work is the relation between the gravitational properties of the solutions and their non-linear nature. In Fig. \ref{pd} we then present the phase diagram of Einstein-Weyl gravity, where we show the families of solutions present for each pair of values of the $M$ and $S_2^-$ parameters of the gravitational potential. 

This diagram is mainly populated by type I, type II and no-sy WHs, while black holes are located on the 1-dimensional region between wormholes and type II solutions, that corresponds to the Schwarzschild line $S_2^-=0, M>0$, and to the non-Schwarzschild black hole curve. Being on a zero-measure region, the first thing emerging from this diagram is that black holes are not expected to be the general vacuum solutions of Einstein-Weyl gravity, unless specific arguments are taken into account. Stressing the phase diagram analogy, we can interpret them instead as a transition between type II solutions and no-sy WHs.

The second important feature shown in Fig. \ref{pd} is that the positive mass region is mainly populated by type I and wormhole solutions. We note that, for arbitrary large mass values, the phase diagram suggests that type I, type II and no-sy WHs appear, so it can be expected that such solutions can appear also with astrophysical mass values. However, while no-sy WHs are present for any negative value of the Yukawa charge in this limit, for a positive value there is a qualitative difference in having very small values of $S_2^-$, which would lead to a type II solution, or a larger value, which instead implies the presence of a type I solution. Regarding non-Schwarzschild black holes, instead, they are present in the positive mass region only for limited values of the ADM mass, of the Yukawa charge, and of the horizon radius, suggesting that they can be relevant only at microscopic scales. In the small mass region in fact we find a different configuration in which type II solutions are present for a negative Yukawa charge, and no-sy WHs for positive Yukawa charge. We also mention that in this region appears the unique symmetric and asymptotically flat wormhole in both side of the spacetime of the theory at the point $M\simeq0.61, S_2^-\simeq0.11$, which is, aside the minkowski spacetime, the unique vacuum solution with regular curvature invariants in all the available space by an observer.        

We find all the three types of solutions also in the negative mass region, where there is a dominance of type I solutions, but also a greater presence of type II solutions with respect to the positive mass region. For a positive value of the Yukawa charge type I solutions are the only family present in the phase diagram, while for negative values of $S_2^-$ there is a relevant presence of type II for smaller absolute values of the mass, and of no-sy WHs for even smaller values. It has to be noted, however, that the line that delimits the transition between type I and type II is sensitive to the specific value of the radius in which we choose to evaluate the functions (\ref{chis}), and then has to be taken with care.

\begin{figure}[ht]
\centering
\includegraphics[width=\columnwidth]{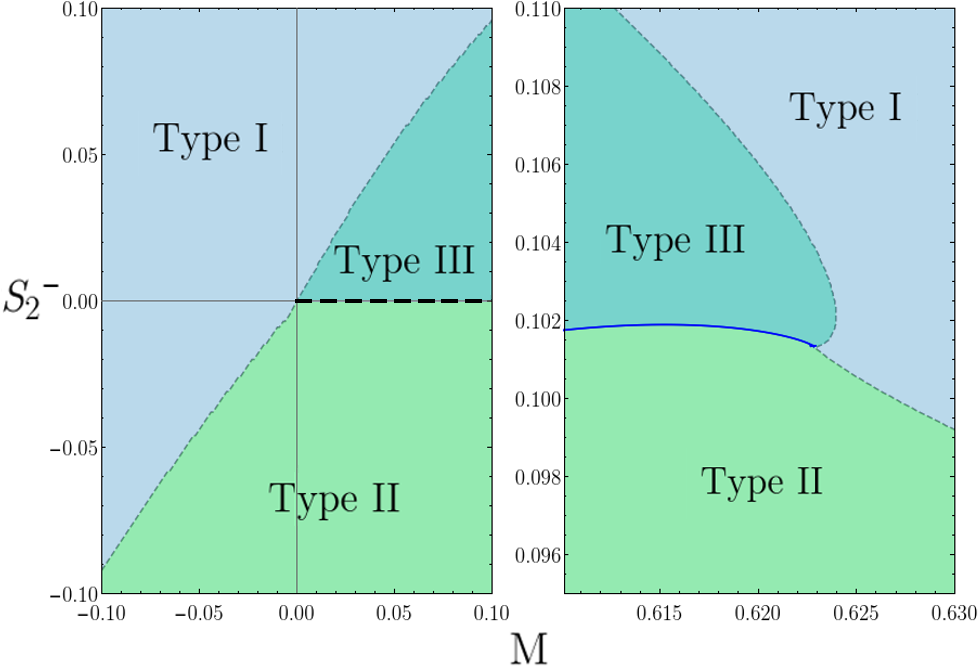}
\caption{The two triple points of the phase diagram; the first one (Minkowski triple point) is located at the origin of the $M$-$S_2^-$ plane, while the second one (massive triple point) is located at $M\simeq0.623,\, S_2^-\simeq0.102$, that is the endpoint of the non-Schwarzschild black hole line.}
\label{triplepoint}
\end{figure}

Stressing even more the phase diagram analogy, we see that at the border of the transition line between type I solutions and no-sy WHs, two triple points appear; we higlight these two points in Fig. \ref{triplepoint}. The first triple point corresponds to the Minkowski spacetime $M=0, S_2^-=0$, while the second one, which we will call massive triple point from now on, is located at $M\simeq0.623,\, S_2^-\simeq0.102$ in our numerical units. Both triple points are at the border of the areas populated by type I, type II, black holes and wormholes. Moreover, the Minkowski one corresponds to the point where the horizon of Schwarzschild black holes goes to zero, while the massive triple point is where the horizon of non-Schwarzschild black holes goes to zero. This fact corresponds also to a vanishing value for various quasi-local masses definition evaluated at the horizon, \textit{i.e.} a vanishing value of the energy enclosed inside the horizon. Taking into account these properties, it seems that the Minkowski flat space is not unique as it is in General Relativity, and could not be the only ``true vacuum'' of the theory; the presence of this massive triple point suggests that there might be a sort of ghost condensate vacuum, that might have relevant consequences on the study of quantum fluctuations.

\subsection{The phase diagram in presence of matter}

\begin{figure}[ht]
\centering
\includegraphics[width=\columnwidth]{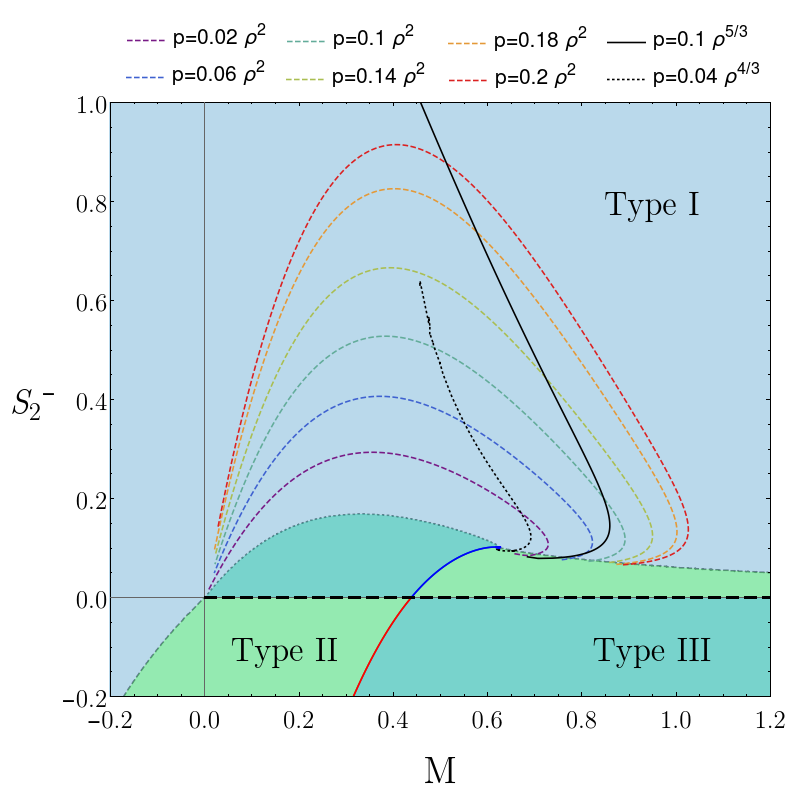}
\caption{Location of non-vacuum solutions in Einstein-Weyl gravity; the dashed lines indicate solutions with polytropic equations of state with $\Gamma=2$, with different colors for different values of $k_0$, while the solid and dotted black lines indicate solutions with an equation of state with $\Gamma=5/3$ and $\Gamma=4/3$ respectively.}
\label{pdtov}
\end{figure}

The location of non-vacuum solutions in the phase diagram is shown in Fig. \ref{pdtov}. The change of the parameter $k_0$ in the e.o.s. (\ref{poly}) results in a scale transformation in the parameter space, as expected by its relation to the energy scales of the fluid, while the change in the polytropic exponent $\Gamma$ modifies the behaviour of the $M$-$S_2^-$ relation. Nonetheless there are some important common features:
\begin{itemize}
\item[-] the non-vacuum solutions are present only in the area populated by Type I solutions;
\item[-] as the star radius decreases, the solutions converge to the massive triple point of the phase diagram.
\end{itemize}
The main qualitative difference is that the solutions with $\Gamma=2$, for which a vanishing central pressure is reached at a finite radius and with vanishing mass, converge also to the Minkowski triple point, while the solutions with $\Gamma=4/3,\,5/3$, for which a vanishing central pressure instead is reached in the large star radius limit and with a constant mass, seem to have a divergent value of the Yukawa charge. We have to note, however, that for very large star radii the integration becomes unstable, due to the fact that the weak field limit is not reliable anymore.

The two common features, however, highlight the main physical aspects of Fig. \ref{pdtov}. The location of the non-vacuum solution in the phase diagram sheds new light on Type I solutions, that did not had a physical interpretation until now. They appear in fact as the external field of compact objects, and therefore are a candidate to be the generic observed solution in a quadratic theory of gravity. We note that there is an indication that this might not be true in the full quadratic theory \cite{Silveravalle:2022lid} but, as said in the introduction, the complete integration of such theory is still not completely understood.
The second aspect is that the two triple points of vacuum solutions play a special role also for non-vacuum solutions. While the Minkowski flat space remains an attractive point for solutions which vanish for vanishing energy density (property which is not true for all the equations of state also in General Relativity), the massive triple point is an attractor for stars with a divergent central energy density independently of the equations of state of the fluid. While it is still not clear the role of this massive triple point, we believe that there is much evidence that indicates that it might be relevant and should be studied with care.

\subsection{Phenomenological consequences}

\begin{figure*}[t]
\centering
\includegraphics[width=\textwidth]{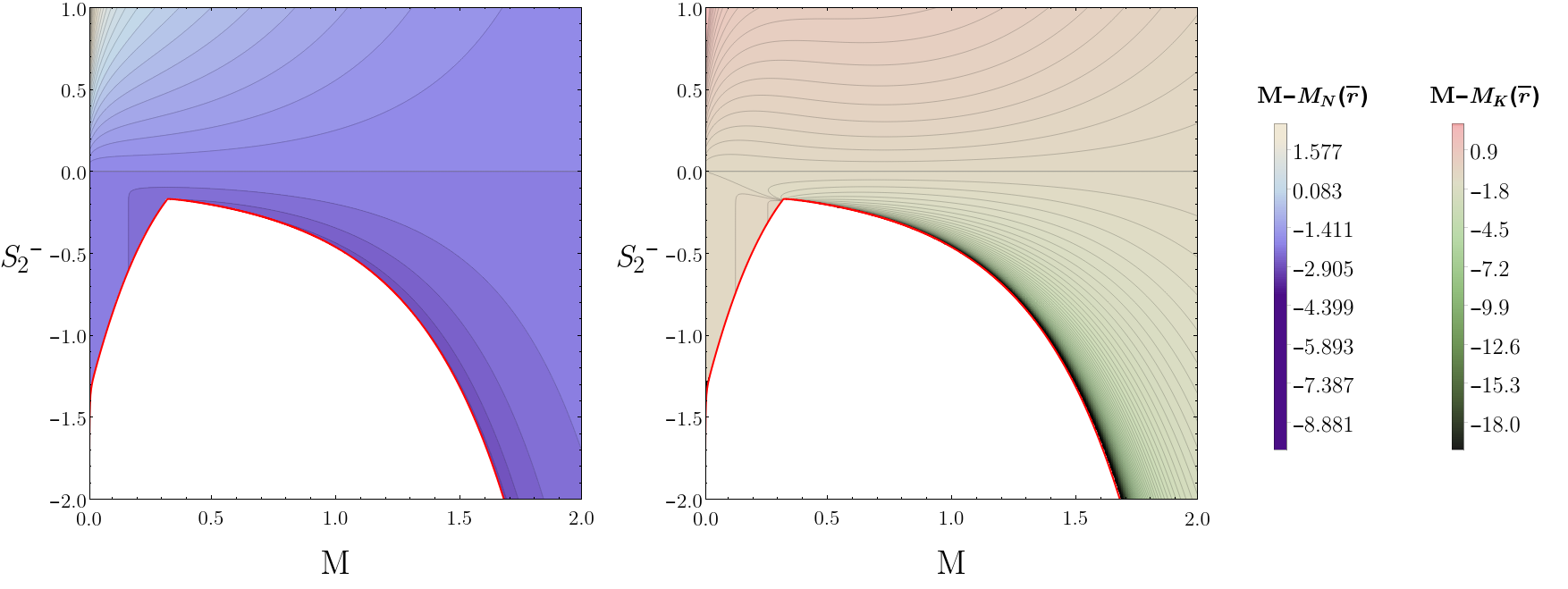}
\caption{Differences in observable masses measured at infinity and at $\Bar{r}=3M$; in the left panel the mass $M_N$ is measured by the redshift of a photon, while in the right panel the mass $M_K$ is measured using Kepler's third law. The white region is removed, being populated by wormholes with throat radius $r_T>3M$.}
\label{fig5}
\end{figure*}

As already stated before, the main advantage of having pictured the phase diagram of the theory is to have a direct link between the precise form of the gravitational field and the small scale nature of the solutions. However, an explicit measure of the strength of the Yukawa term might be problematic, and it would also be strongly model dependent. Nonetheless is possible to approach the phenomenology of quadratic gravity in a different way, that is to model the discrepancies in the results of different experiments made using standard techniques of General Relativity, as discussed in \cite{Bonanno:2021zoy}.

As a toy-model example we can imagine to measure the mass of an object with the redshift of a photon emitted at a radius $\Bar{r}$ by some gas using
\begin{equation}\label{massz}
M_N(\Bar{r})=\frac{1}{2}r\left(1-h(\Bar{r})\right)=\frac{1}{2}r\left(1-\frac{1}{\left(1+z(\Bar{r})\right)^2}\right),
\end{equation}
where $z$ is the redshift, or by the transit of a satellite, assuming that in some limit we can use Kepler's third law, using
\begin{equation}\label{massk}
M_K(\Bar{r})=\frac{1}{2}r^2h'(r)=\left(\frac{2\pi}{T}\right)^2\Bar{r}^3,
\end{equation}
where $T$ is the orbital period. Both measurements will coincide with the ADM mass parameter $M$ at infinity, but will have discrepancies when evaluated at finite radii. In Fig. \ref{fig5} we show how these mass definitions differ from the asymptotic value if measured at a radius $\Bar{r}=3M$, which is the radius of the photon sphere for a Schwarzschild black hole; we have removed the area of the phase diagram populated by no-sy WHs with throat radius greater than $3M$ which, however, is not relevant in the large mass limit. As could be expected, the discrepancy increases for small masses but is still present at any scale. In particular, in the limit in which the spacetime at $\Bar{r}$ is already in the linearized regime, the differences become
\begin{equation}\label{massdif}
\begin{split}
M-M_N(\Bar{r})&=S_2^-\mathrm{e}^{-m_2\, \Bar{r}},\\
M-M_K(\Bar{r})&=S_2^-\mathrm{e}^{-m_2\, \Bar{r}}\left(1+m_2\,\Bar{r}\right).
\end{split}
\end{equation}
While both measurements will be exponentially suppressed, is still noticeable the linear dependence from the Yukawa charge $S_2^-$, which means that an experiment which measure the mass close to the object will see a larger mass for negative $S_2^-$ and a smaller one for positive $S_2^-$. In the large mass limit, this directly translate in the presence of a no-sy WH or of a type I naked singularity.

Similar arguments where made in the context of compact star phenomenology in \cite{Bonanno:2021zoy,Silveravalle:2022lid}, in which is discussed that stars with different equations of state will have different discrepancies in the mass definitions. This argument could be of great interest for the study of neutron stars, opening the possibility of discriminating different equations of state using only gravitational measurements, and without having to evaluate the surface radius of the star.

\section{Conclusions}

In this paper we give a complete description of the solution space of Einstein-Weyl gravity. We use the analytical approximation of the solutions known, which at large radius are given by a Newtonian potential plus a Yukawa massive correction characterized by the ghost mass of the theory, while at finite radius are given in series form. We connect the finite and large radius regimes by means of numerical integrations together with a shooting method in case of black hole and wormhole solutions. We review the main type of vacuum solution found in previous work  and we show the behavior of their gravitational potential and their causal structure in relation with their position on the solution space. We present the solution space of the theory in the form of phase diagram, in which the type of vacuum solution is shown in function of the asymptotic parameter, i.e. the mass and the Yukawa charge. The phase diagram shows that for an arbitrary positive large mass we find: Schwarzschild black holes confined on the one-dimensional region with zero Yukawa charge, non-symmetric wormholes on the two dimensional region with negative Yukawa charge, type II solution (attractive naked singularity) for small positive Yukawa charge and type I solutions (repulsive naked singularity) for large positive Yukawa charge. Therefore we expect that all these types of solution can appear also with astrophysical size, showing that the static spherically symmetric vacuum solutions of quadratic gravity admit a large variety of candidates together with black holes. On the other hand, the non-Schwarzschild black hole curve appears only in the small mass region and in the negative mass region, suggesting that such solutions can appear only with microscopical size.
The previous considerations open up the question of what is the final state of a collapsing mechanism in quadratic gravity since, without any further assumption, it is reasonable to consider the solution families covering two-dimensional regions as more available candidates than black holes. 
In the small mass region, which corresponds to solutions with a mass value comparable with the mass of the ghost, we find a different configuration of the types of solution, and two triple points appear. The first one corresponds to the Minkowski space-time
while the second one can be interpreted as a second vacuum of the theory constituted by a ghost condesate. 
The coupling with a matter source is studied by considering a perfect fluid stress-energy tensor with a polytropic equation of state. The resulting gravitational field always has positive mass and positive Yukawa charge, always corresponding to the field of Type I vacuum solutions. Recalling that the type I vacuum family has an attractive field at large distance but repulsive around the origin, when coupling with matter, the resulting field is still attractive but much weaker than their General Relativity counterpart. This gives a different mass-radius relation that makes the same pressure being able to sustain more massive stars. 
The second common feature emerging is that, as the star radius decreases, the solutions converge to the ghost triple point of the phase diagram.
The link between the gravitational potential and the small scale nature of the solutions given by the phase diagram opens the possibility of having a phenomenological signature of quadratic gravity. In particular a mass measurement made close to an isolated object will have a different value than a measurement made at infinity, and thanks to the phase diagram we could use such differences to made a prediction on the physical nature of the solution. For compact stars, such prediction would be on the equations of state of the fluid.
These considerations add new information about the classical solution of Einstein-Weyl gravity and their physical relevance. In order to make a further step in this direction, we will consider in future the stability of the solutions and the result of a collapsing mechanism, which are crucial to understand what kinds of astrophysical objects are predicted by quadratic gravity together with (or in substitution of) black holes.     

\acknowledgements 

The authors would like to thank Alfio Bonanno for the constant support, for his useful advices and the stimulating discussions.

\bibliographystyle{apsrev4-1}

\end{document}